# A new standard quadratic optimization approach to beam angle optimization for fixed-field intensity modulated radiation therapy

Junbo Peng[1], Ailin Wu[2] and Lei Zhu[1]
[1]School of Physical Sciences, University of Science and Technology of China, Hefei, Anhui 230026, China
[2]The First Affiliated Hospital, University of Science and Technology of China, Hefei, Anhui 230001, China

**Abstract:** Beam angle optimization (BAO) largely determines the performance of fixed-field intensity modulated radiation therapy (IMRT), and it is usually considered as non-convex optimization and an NP hard problem. In this work, BAO is reformulated into a highly efficient framework of standard quadratic optimization. The maximum of beamlet intensities for each incident field as the surrogate variable indicating whether one radiation field has been selected. By converting the function of maximum value in the objective into a set of linear constraints, the problem is solved as standard quadratic optimization via reweighting l1-norm scheme. The performance of the proposed BAO has been verified on a digital phantom and two patients. And the conclusion is drawn: the proposed convex optimization framework is able to find an optimal set of beam angles, leading to improved dose sparing on OARs in fixed-field IMRT.

## 1. Introduction

Inverse treatment planning for intensity modulated radiation therapy (IMRT) aims to obtain a prescribed dose distribution on planning target volume (PTV) while sparing organs at risk (OARs). A fully optimized IMRT plan should consider all the system paremeters of a clinical linear accelerators as control variables in the optimization process, including beam number, beam angle, multi-leaf collimator (MLC) leaf positions, and monitor unit (MU) ofr each segment. Convex formulation of such an optimization task, however, appears challenging since most control variables have a non-linear relationship with the delivered dose distribution. In this work, we improve fixed-field IMRT by including beam angled optimization into the inverse treatment planning process, via a new $l_1$-norm minimization approach.

A large number of treatment beams prolongs dose delivery time and therefore increases potential dose errors due to patient motion. On the other hand, it is reported that the dose improvement of a treatment plan diminishes as beam number increases and less than 10 beam angles are often sufficient for IMRT[1]. As a small beam number is used in current fixed-field IMRT, the selection of beam angles largely determines the treatment plan quality[2, 3]. Beam angle optimization (BAO) searches for an optimal set of beam orientations to obtain the best plan quality from all possible beam angle combinations, which is inherently an NP-hard combinatorial optimization problem with no efficient solutions yet[4-6]. As such, BAO is not ubiquitously implemented in current clinical practice. Instead, beam number is first empirically determined, and beam angles are selected in a trial-and-error fashion. Due to the mathematical complexity of inverse planning in IMRT, empirical tuning of beam angle selection does not guarantee the optimality of treatment plan. For instance, the mathematically optimal beam configuration can be counterintuitive since the extra freedom of intensity modulation compensates for the visually sub-optimal beams[7]. To shorten the treatment planning time of IMRT, equiangular beams are used in many radiation therapy scenarios, and the same beam angle setting is typically used for the same disease site on different patients, at the cost of reduced plan optimality[8, 9].

BAO for IMRT has been an active research area for decades[6, 7, 10-14]. Many existing BAO methods improve the empirical selection of beam angles by including dosimetric or geometric considerations[12, 15-17], and they are not exactly optimization algorithms from a mathematical perspective. For example, [12] uses beam's-eye-view dosimetrics to rank the possible beam orientations by evaluating the quality of an achieved dose distribution without exceeding the tolerance of OARs for each beam candidate. Another similar work[16] uses the ratio of OAR total dose to mean PTV dose as the quality metric for each incident field. The above strategies reduce the computation of BAO by analysing the contribution of individual beam to the overall quality of a treatment plan, which inevitably compromises the optimality of delivered dose distribution due to negligence of multiple-beam interplay[18].

Another category of BAO methods aims to find the optimal beam angles for IMRT using global optimization for a non-convex problem. Existing approaches include simulated annealing algorithms[7, 13, 18], genetic algorithms[19, 20], particle swarm optimization method[21], and multi-objective optimization algorithms[22, 23]. As a weakness of non-convex optimization with a large solution pool in general, these methods typically require clinically unacceptable long computation and it is theoretically impossible to guarantee the global optimality of the solution due to the existence of multiple local minima[10, 24].

Recent developments on optimization methods give rise to non-conventional treatment planning algorithms for IMRT. Sparse optimization was introduced to IMRT treatment planning by Zhu and Xing to obtain a satisfactory dose

Corresponding E-mail: leizhusg@ustc.edu.cn

distribution with a simplified treatment plan[25, 26]. By minimizing a total-variation objective with quadratic constraints, the algorithm finds piece-wise constant fluence maps with sparse gradients, leading to a highly efficient treatment with a small number of segments. BAO searches for optimal sparse beams in the angular space, which can be formulated as a sparse recovery problem as well. The key challenge of solving BAO via sparse optimization is to find an appropriate control variable for the objective function to indicate the sparsity of beams while still preserving the convexity of the optimization problem. A probably first attempt of sparse optimization based BAO can be found in a recent literature[27]. The authors find it difficult to formulate an $l_1$-norm objective and propose a mixed $l_{2,1}$-norm of beam intensities instead, which is inherently an adaptation of the group-lasso (also called group-sparsity) method[28] widely used in signal processing and statistical learning[29-33] for fearture selection. Theoretically, such a scheme compromises the optimality of BAO for the mixed $l_{2,1}$-norm objective not only imposes the sparsity on the beam angle level, but also perform an additional smoothing penalty within each field. On the other hand, the authors also admit it that the use of a somewhat complicated group-lasso objective also complicates the computation since the proposed $l_{2,1}$-norm minimization cannot be solved by either linear or quadratic programming. Some BAO algorithms are recently developed based on the group-lasso method and they both replace the $l_1$-norm with the nonconvex function to better approximate the $l_0$-norm while rataining the $l_2$-norm in each field as the control variable to indicate whether this field is selected or not[5, 34]. The quasi $l_0$-norm method (i.e. $l_{2,0}$-norm) eliminates the $l_2$-norm penalization within each field during the BAO process, it still employs a complicated iteration scheme due to the nonideal group-sparsity objective.

In this work, by designing a new control variable in the sparse optimization framework, we propose an improved BAO algorithm with an $l_1$-norm regularized quadratic objective. Since the algorithm is in a standard form of quadratic optimization, it accurately finds the global minimum with high computational efficiency. The method performance is demonstrated on one digital phantom, one prostate patient and one head-and-neck (HN) patient.

## 2. Methods

### *2.1 Inverse treatment planning of IMRT using $l_1$-norm minimization*

We develop the proposed algorithm using a beamlet model. Each radiation beam from a pre-determined angle is divided into small beamlets. The delivered dose distribution on the patient, $\vec{d}$, has a linear relationship with beamlets of fluence map, $\vec{x}$:

$$\vec{d} = A\vec{x} \tag{1}$$

where $\vec{d}$ is a vectorized dose distribution for a three-dimensional volume, and the beamlet intensity $\vec{x}$ is a one-dimensional vector that consists of row-wise concatenations of beamlet intensities for all fields. Each column of the matrix $A$ is a beamlet kernel which corresponds to the delivered dose distribution by one beamlet with unit intensity. In this work, we use the Voxel-based Monte Carlo algorithm (VMC)[35] to generate the matrix $A$.

In the conventional beamlet-based treatment planning of IMRT, sum of square errors of the delivered dose relative to the prescribed dose is used as an objective function in the optimization of the beamlet intensity $\vec{x}$, and the fluence map optimization (FMO) problem is expressed as:

$$minimize\text{: } \phi_{FMO} = \sum_i \lambda_i \, (A_i\vec{x} - d_i)^T (A_i\vec{x} - d_i) \tag{2}$$
$$subject\ to\text{:} \quad \vec{x} \geqslant 0$$

where the index $i$ denotes PTV or different OARs, $A_i$ is the beamlet kernel for different structures, $\lambda_i$ is the corresponding importance factor[36, 37], and $d_i$ is the prescribed dose to each structure. The optimized beamlet intensity is finally converted to MLC leaf positions and MUs for different segments, using a leaf sequencing algorithm[38].

In current fixed-field IMRT, a small number of beam angles (typically 5-10) are pre-determined before the optimization of beamlet intensities. In this work, we aim to include a large number of beam angles from a full rotation into the beamlet optimization framework and use sparse optimization to automatically select the optimal beam combination, so the new optimization algorithm takes the following form of $l_1$-norm minimization:

$$minimize\text{:} \quad \phi_{FMO} + \beta \left\| \vec{S}(\vec{x}) \right\|_1 \tag{3}$$
$$subject\ to\text{:} \quad \vec{x} \geqslant 0$$

where $\beta$ is a user-defined parameter adjusting relative weights between the FMO objective and the BAO objective; $\vec{S}(\vec{x})$ is a vector with a length of the total available beam number, and one element of $\vec{S}(\vec{x})$ is zero if the corresponding beam angle is not selected. $\|\cdot\|_1$ calculates the $l_1$ norm of one vector. Note that, the beamlet kernel $A_i$ in the optimization problem (3) has a significantly increased size as compared with that in the optimization problem (2), due to the large number of beamlets from all available fields.

The function $\vec{S}$ in the optimization problem (3) outputs a sparse vector signal when only a small number of beam angles are selected and the combined objective favors a sparse $\vec{S}$ due to the existence of $l_1$-norm regurlarization term. The design of $\vec{S}$ is the main contribution of this paper. The challenge lies in that the optimization problem (3) needs to be in a form of or convertible to convex optimization for its efficient computation. We propose to use:

$$\vec{S}(\vec{x}) = \max\,(\vec{x}_\theta) \tag{4}$$

where $\vec{x}_\theta$ denotes all beamlets at angle $\theta$, and max $(\vec{x}_\theta)$ is a vector with a length of total available beam number, of which each element is the maximum intensity of beamlets within one beam at angle $\theta$.

The function of Eq. (4) is non-linear or quadratic. However, it can be easily verified that the proposed optimization framework, i.e. the optimization problem (3) and Eq. (4), can be converted to an equivalent form of quadratic optimization. Define a new vector $\vec{y}$, with a length of total available beam number. The proposed algorithm can be rewritten as:

$$\begin{aligned} &minimize: \quad \phi_{FMO} + \|\vec{y}\|_1 \\ &subject\ to: \quad \vec{x} \succcurlyeq 0, \vec{x}_\theta \preccurlyeq \vec{y}(\theta) \end{aligned} \tag{5}$$

where $\vec{y}(\theta)$ stands for the element of vector $\vec{y}$ at angle $\theta$. Compared with the mixed $l_{2,1}$-norm and $l_{2,0}$-norm methods mentioned before[5, 27, 34], our algorithm exhibits the form of $l_{2,\infty}$-norm which excludes the smoothing penalization within each field during the BAO process without sacrifacing the computation efficiency of convex optimization.

In order to convert the problem (5) to the framework of a standard quadratic programming, we define a combined vector $\vec{z} = [\vec{x}; \vec{y}]$. If the total number of beamlets is N and the number of all candidate beam angles is M, so the optimization variable $\vec{z}$ is an (N+M)-dimensional vecotr. Then the problem (5) could be rewritten as an equivalent form as:

$$\begin{aligned} &minimize: \quad \frac{1}{2}\vec{z}^T Q \vec{z} + \vec{c}^T \vec{z} \\ &subject\ to: \quad \mathrm{B}\vec{z} \preccurlyeq 0 \end{aligned} \tag{6}$$

where Q is an $(N+M)^2$-dimensional real symmetric matrix, $\vec{c}$ is a vector with the length of (N+M), and B is an $(N+M)^2$-dimensional matrix. Specifically,

$$Q = \begin{bmatrix} 2A^T A & O \\ O & O \end{bmatrix} \tag{7.1}$$

$$\vec{c} = \begin{bmatrix} -2A^T \vec{b} \\ \beta \cdot \vec{e} \end{bmatrix} \tag{7.2}$$

where $\vec{e}$ is an M-dimensional all-one vector.

$$\mathrm{B} = \begin{bmatrix} -I & O \\ I & \mathrm{B}_0 \end{bmatrix} \tag{7.3}$$

where $I$ is an identity matrix with the size of N by N, and $\mathrm{B}_0$ is expressed as

$$\mathrm{B}_0 = \begin{bmatrix} -1 & 0 & 0 & \cdots \\ -1 & 0 & 0 & \cdots \\ \vdots & \vdots & \vdots & \vdots \\ 0 & -1 & 0 & \cdots \\ 0 & -1 & 0 & \cdots \\ \vdots & \vdots & \vdots & \vdots \\ 0 & 0 & \cdots & -1 \\ 0 & 0 & \cdots & -1 \\ \vdots & \vdots & \vdots & \vdots \end{bmatrix} \tag{7.4}$$

Multiplication by the upper half of matrix B is used to perform the constraints of $\vec{x} \succcurlyeq 0$, while multiplication by the other half of matrix B imposes the contrasints of $\vec{x}_\theta \preccurlyeq \vec{y}(\theta)$ on the optimization variable.

The problem (6) has a form of standard quadratic optimization, and it is the main result of the paper. Zero elements of the optimized $\vec{y}$ obtained from the optimization problem (6) indicates that the corresponding beam angles should not be used in the fixed-field IMRT.

## *2.2 The proposed BAO with a reweighting scheme*

Derived from $l_1$-norm minimization, the optimization problem (6) sacrifices sparsity of the optimized solution for computational efficiency. At the cost of increased computation, the non-convex $l_0$-norm minimization enhances the solution sparsity and therefore reduces the number of required beams. In this paper, we propose to balance the computational efficiency and the solution sparsity via a series of reweighted $l_1$-norm minimization, a strategy commonly used in different $l_1$ based sparse optimization problems[27, 39, 40].

The reweighting scheme approximates $l_0$-norm minimization by adaptively assigning large weights to the optimized vector elements with small values in the previous iteration of $l_1$-norm minimization[40]. In each iteration, the optimization takes the following form:

$$\begin{aligned} &minimize: \quad \frac{1}{2}\vec{z}^T Q \vec{z} + \overrightarrow{c'}^T \vec{z} \\ &subject\ to: \quad \mathrm{B}\vec{z} \preccurlyeq 0 \end{aligned} \tag{8}$$

where $\overrightarrow{c'}$ is defined as $[-2A^T\vec{b}\,;\ \beta \cdot \vec{w}]$. The optimization problem above has the same form of problem (6), except that the all-one vector $\vec{e}$ is replaced by a new weighting vector $\vec{w}$. $\vec{w}(\theta)$, element of $\vec{w}$ at beam angle $\theta$, is adpatively updated after each iteration using the same methods as in[27]:

$$\vec{w}(\theta) = exp\left[1 - \frac{\|x_\theta\|_2}{\|x_\theta\|_2^{max}}\right] \tag{9}$$

where $\|x_\theta\|_2$ calculates the $l_2$-norm of all beam intensities at angle $\theta$ and $\|x_\theta\|_2^{max}$ is the maximum of $\|x_\theta\|_2$ in the neighboring three angles with non-vanishing values.

**Algorithm:** BAO using quadratic optimization with reweighting

Set the parameter values of $\lambda_i$ , $d_i$ and $\beta$; Initialize $\vec{w}(\theta) = 1$ for all $\theta$. Set a target beam number $N_A$.

**repeat**

1. Solve the optimization problem (8);
2. Count the number of non-zero elements in $\vec{y}$, $N_{ang}$;
3. Update $\vec{w}$ using Eq. (9).

**until** $N_{ang}$ is no longer larger than $N_A$.

The proposed BAO algorithm is summarized above. We first initialize $\vec{w}$ as an all-one vector. The optimization problem (8) is repeatedly computed with $\vec{w}$ updated using Eq. (9). After each iteration, we count the number of non-zero elements of $\vec{y}$, i.e., the number of selected beam angles, $N_{ang}$. The BAO process terminates if $N_{ang}$ is no longer than $N_A$. After

BAO selects the optimal beam angles, a standard inverse planning for fixed-field IMRT finally generates a treatment plan and a delivered dose distribution.

### *2.2 The proposed BAO with a reweighting scheme*

We evaluate the proposed BAO method on a digital phantom, a prostate patient and a head-and-neck patient. For all evaluation studies, we consider 40 equiangular beams in a full rotation as the candidates of all available beam orientations. The PTV is centered at the axis of rotation, with a source-to-axis distance (SAD) of 100 cm. In the FMO for fixed-field IMRT, each field targets the center of PTV, and contains 20 by 20 beamlets, with a beamlet size of 5 mm by 5 mm at SAD, while during the BAO process, the beamlet size is downsampled to 1cm by 1cm to reduce the memory usage. To save computation in the Monte Carlo simulation of the dose kernel (i.e., the matrix $A_i$), the CT data are downsampled to a voxel size of 3.92 mm by 3.92 mm by 2 mm. All the algorithms are implemented in Matlab, using CVX, an open-source optimization software[41]. On a 2.4 GHz workstation with 28 cores, the proposed BAO takes 3 minutes on the digital phantom, 5 minutes on the prostate patient, and 6 minutes on the head and neck patient.

A theoretically optimal set of beam angles is difficult to derive on clinical cases, since it is dependent on the geometries of structures (i.e., the dose kernel $A_i$) as well as the parameters of treatment planning (i.e., $\lambda_i$ , $d_i$ and $\beta$). The study of digital phantom with a known optimal set of beam angles is designed to test the proposed BAO algorithm. We implement the conventional IMRT planning (i.e., the optimization problem (2)) with all beam angles included for comparison. In the patient studies, we investigate the dose performance of fixed-field IMRT using the proposed BAO and a set of equiangular beam angles. In addition to the final dose distributions, we compare the dose-volume-histogram (DVH) curves of OAR for different plans with a similar dose coverage on PTV.

A particular difficulty occurring in the design of patient studies is that, on the same patient, the parameters of IMRT planning, especially the importance factors (i.e., $\lambda_i$), need to be fine-tuned for different sets of beam angles to achieve clinically acceptable treatment plans, leading to unfair comparisons of different algorithms. For a comprehensive evaluation of method performance, we consider fixed-field IMRT planning with the proposed BAO a multi-objective optimization problem, with the following objectives of minimization:

*PTV dose objective:*

$$\phi_1 = (A_{PTV}\vec{x} - d_{PTV})^T(A_{PTV}\vec{x} - d_{PTV})$$

*OAR dose objective*:

$$\phi_2 = \sum_{i \in OARs} \lambda_i \, (A_i\vec{x} - d_i)^T(A_i\vec{x} - d_i)$$

*beam number objective*:

$$\phi_3 = \|\vec{y}\|_1$$

Definitions of all variables are the same as those in the optimization problem (8). The regularization term in problem (5) , $\phi_3$, is considered as one of the objectives in this multi-objective optimization problem, so the penalty weight $\beta$ is no longer needed here and the weighting factors for each angle $\overline{w}(\theta)$ is adpatively adjusted as mentioned before. Note that, for simplicity of result presentation, we use one single dose objective for all OAR structures. The relative values of importance factors of the OAR structures remain unchanged in each patient study. In the evaluations, we fix the value of one objective and compute the Pareto frontiers of the other two objectives.

## 3. Results

### *2.1 The digital phantom study*

Figure 1 (a) shows the water cylinder phantom used in the simulation. The phantom is rotationally symmetric, and therefore the optimal beam angles are only dependent on the relative positions of PTV and OAR. Six passages at randomly selected angles of 0°, 54°, 81°, 153°, 216°, 315° are designed, on which the radiation beams reach PTV without passing through OAR. As such, these six beam angles are considered the optimal orientations in this study.

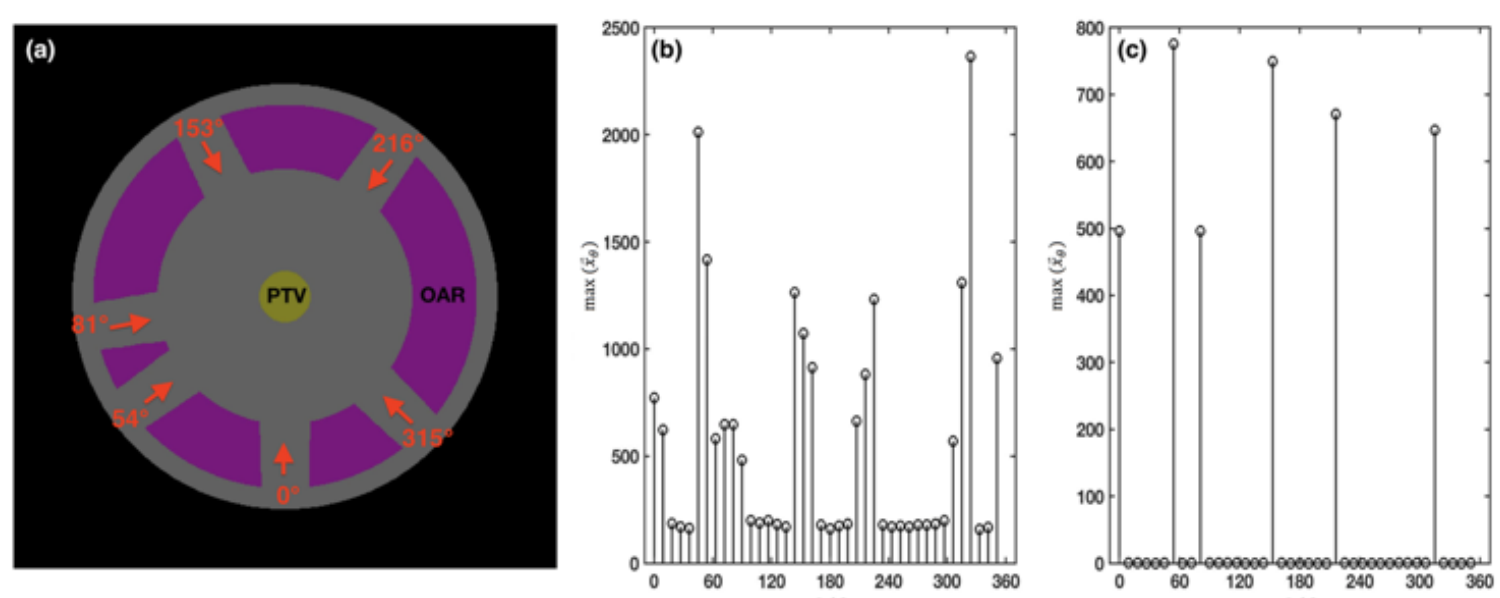

FIG. 1: The validation study on a simulation phantom. (a) The simulation phantom with PTV and OAR. The quantity of $max(\vec{x}_\theta)$ for each incident field without (b) and with (c) the proposed BAO.

The results of conventional IMRT planning using the optimization framework (2) and the proposed BAO algorithm are shown in Fig. 1 (b) and (c), respectively. The maximum value of beamlet intensities for each angle (i.e., max $(\vec{x}_\theta)$ as defined in Eq. (4)) is used as an indicator of whether one beam angle is selected or not. It is seen that the conventional IMRT planning fails to select the most effective beam angles and all 40 beams are used for treatment. The proposed BAO method perfectly chooses the six optimal beam angles with no errors, out of more than 3, 000, 000 possible combinations (i.e., $C_{40}^{6}$).

## *2.2 The prostate patient study*

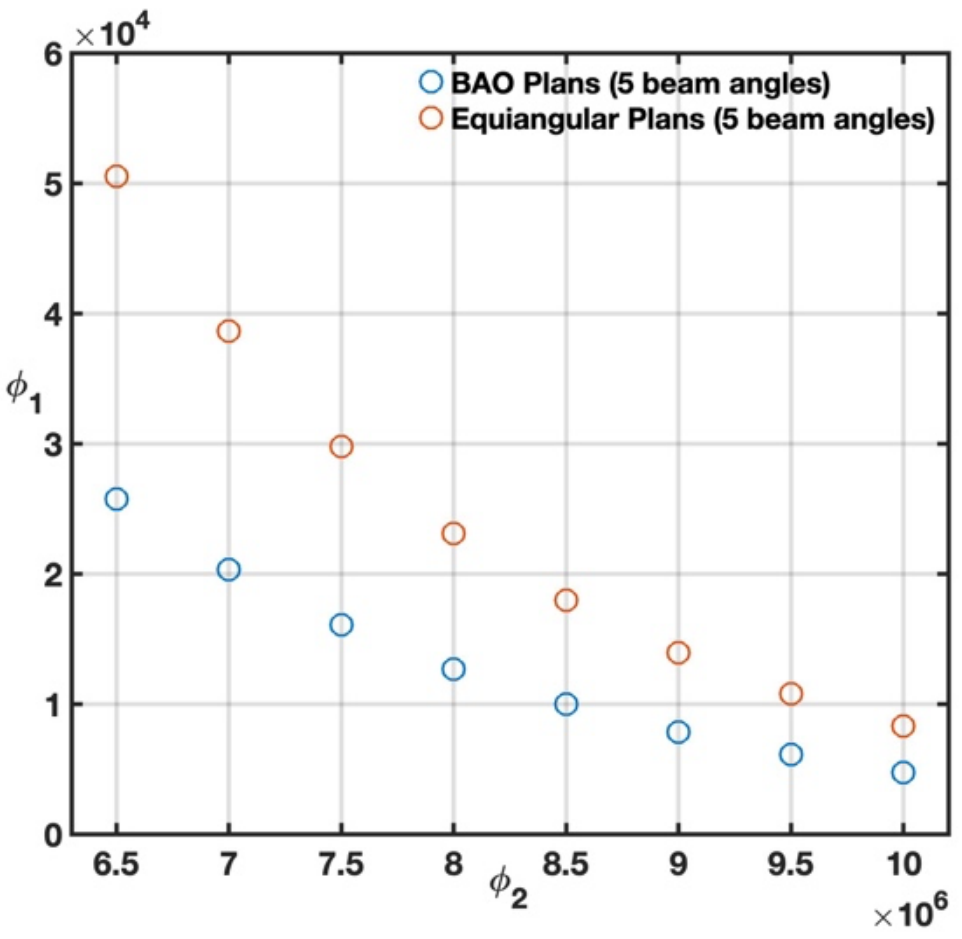

FIG. 2: Results of the prostate patient study. Pareto frontiers of the PTV dose objective ($\phi_1$) and the OAR dose objective ($\phi_2$) for the same number of beam angles using an equiangular plan and the proposed BAO.

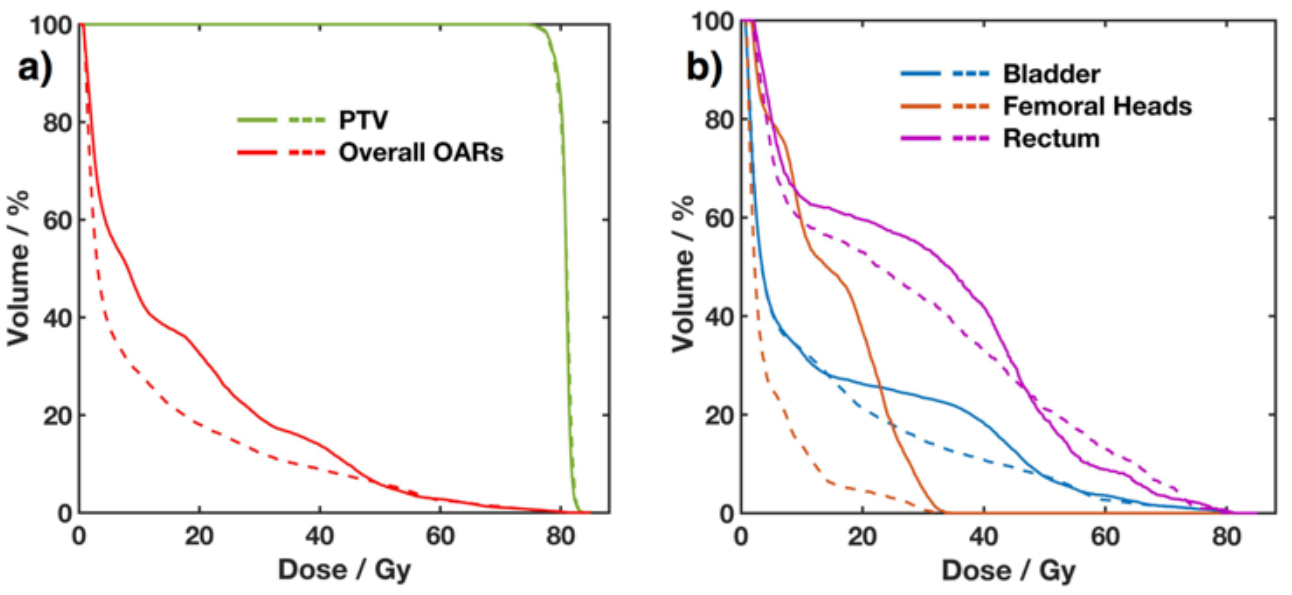

Figs. 2, 3, and 4 show the results on the prostate patient. By tuning algorithm parameters, the proposed BAO is able to select different numbers of beam angles. With the same PTV dose coverage, Fig. 2 compares the Pareto frontiers of fixed-field IMRT with five beam angles using an equiangular plan and the BAO method. It is seen that the proposed BAO substantially improves the dose performance over an equiangular plan with reduced dose objective values on both PTV and OARs.

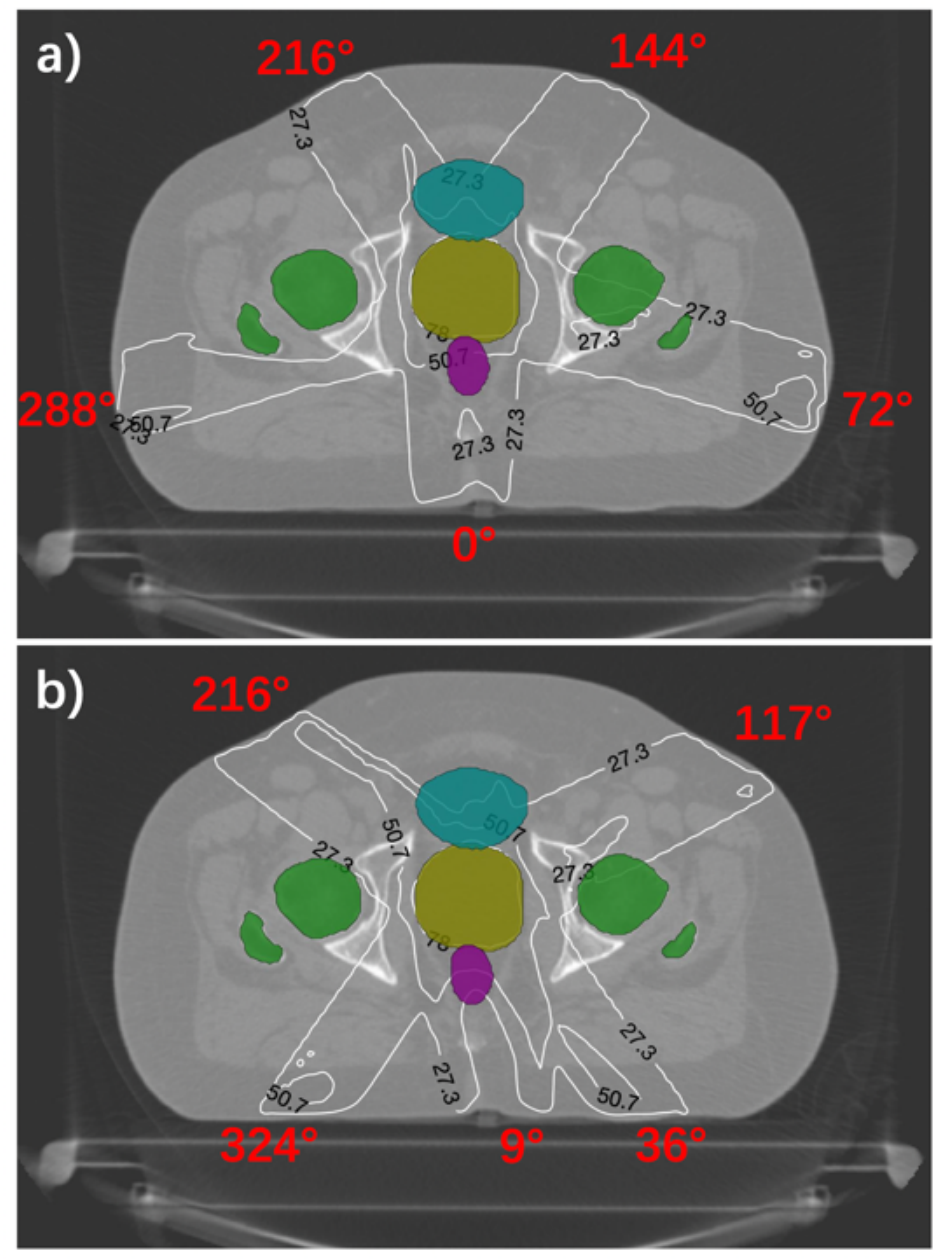

FIG. 3: Comparison of dose distributions on the prostate patient with five beam angles (a) using an equiangular plan and (b) using the proposed BAO, respectively.

FIG. 4: Comparison of DVH curves for the BAO plan (dashed) and the equiangular plan (solid). (a) shows the DVHs in PTV and overall OAR, and (b) shows the DVH in each critical structure.

The improved dose sparing on OARs achieved by the proposed BAO is better seen in the comparison of dose distributions in Fig. 3 and DVHs in Figs. 4. In this patient case, we find that five beam angles (9°, 36°, 117°, 234°, 324°) obtained by the proposed BAO successfully achieve a clinically acceptable dose coverage on PTV. We then compare with the conventional IMRT planning using five equiangular beams starting at 0°. The algorithm parameters are tuned such that both plans obtain the same dose performance on PTV. With the freedom of beam angle selection, the proposed BAO favors beam passages reaching PTV without intersecting OARs (see Fig. 3) and therefore significantly improves dose sparing on OARs. The superior performance of the BAO plan over the equiangular plan is further seen in the DVH comparison of Fig. 4. The proposed BAO reduces the overall dose exposure on OARs by 30.53%. Table 1 summarizes and compares the results of the BAO and equiangular treatment plans with the clinical acceptance criteria, which also demonstrates the validity of the porposed BAO.

Table 1. Prostate plan objectives and results.

| Regions | Acceptance criteria | BAO | Equiangular |
|---|---|---|---|
| PTV | %vol>78 Gy⩾95 | 78.4 Gy | 78.5 Gy |
| Rectum | %vol>70 Gy⩽20 | 51.9 Gy | 49.7 Gy |
| | %vol>60 Gy⩽35 | 38.3 Gy | 42.9 Gy |
| Bladder | %vol>65 Gy⩽25 | 16.8 Gy | 24.9 Gy |
| | %vol>40 Gy⩽50 | 3.4 Gy | 3.5 Gy |
| Femoral Heads | | | |
| Left | %vol>45 Gy⩽1 | 30.9 Gy | 31.8 Gy |
| Right | %vol>45 Gy⩽1 | 16.4 Gy | 32.7 Gy |

## 2.3 The head-and-neck patient study

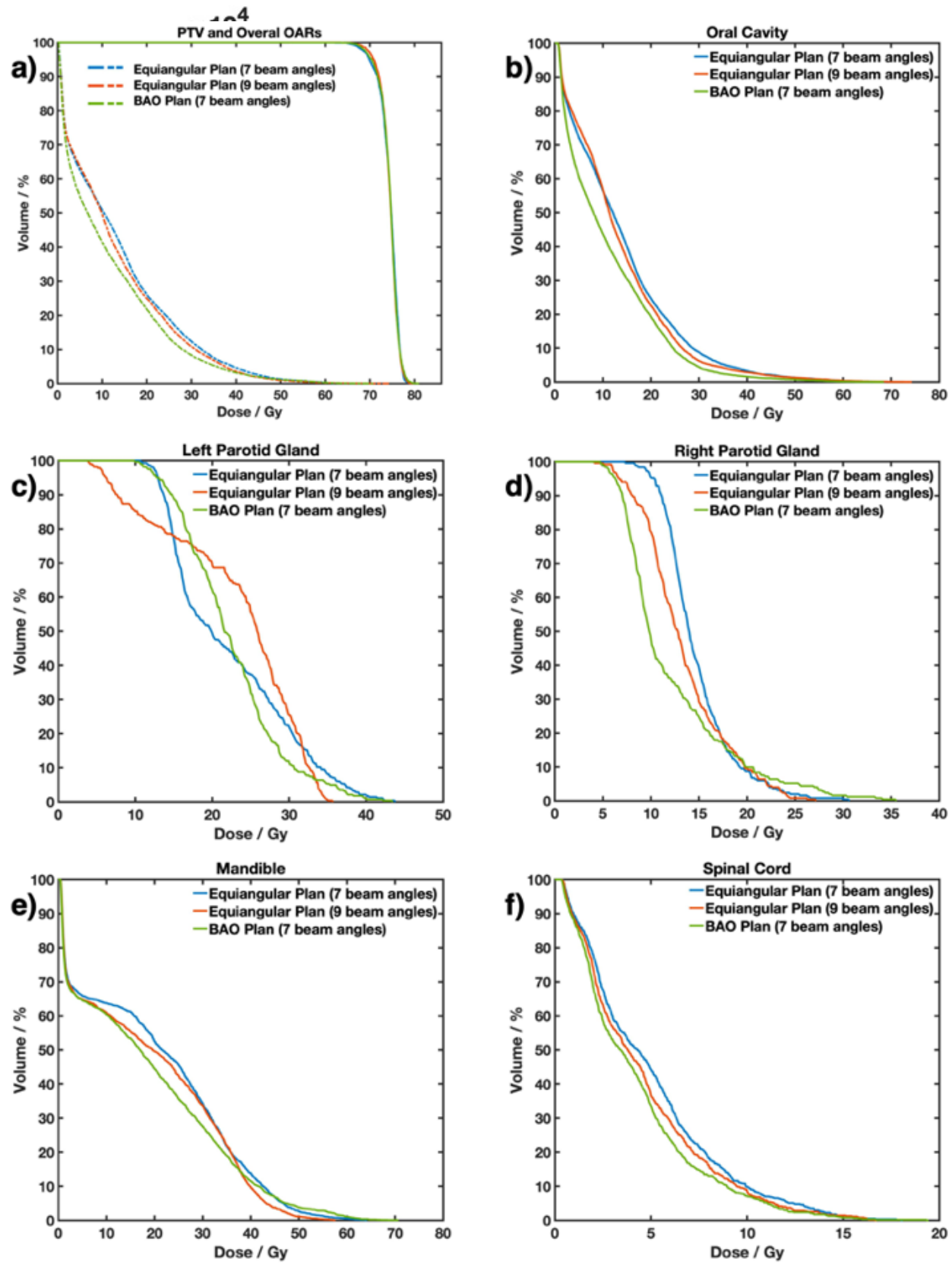

FIG. 5: Results of the head-and-neck patient study. Pareto frontiers of the PTV dose objective ($\phi_1$) and the OAR dose objective ($\phi_2$) for the same number of beam angles using an equiangular plan and the proposed BAO.

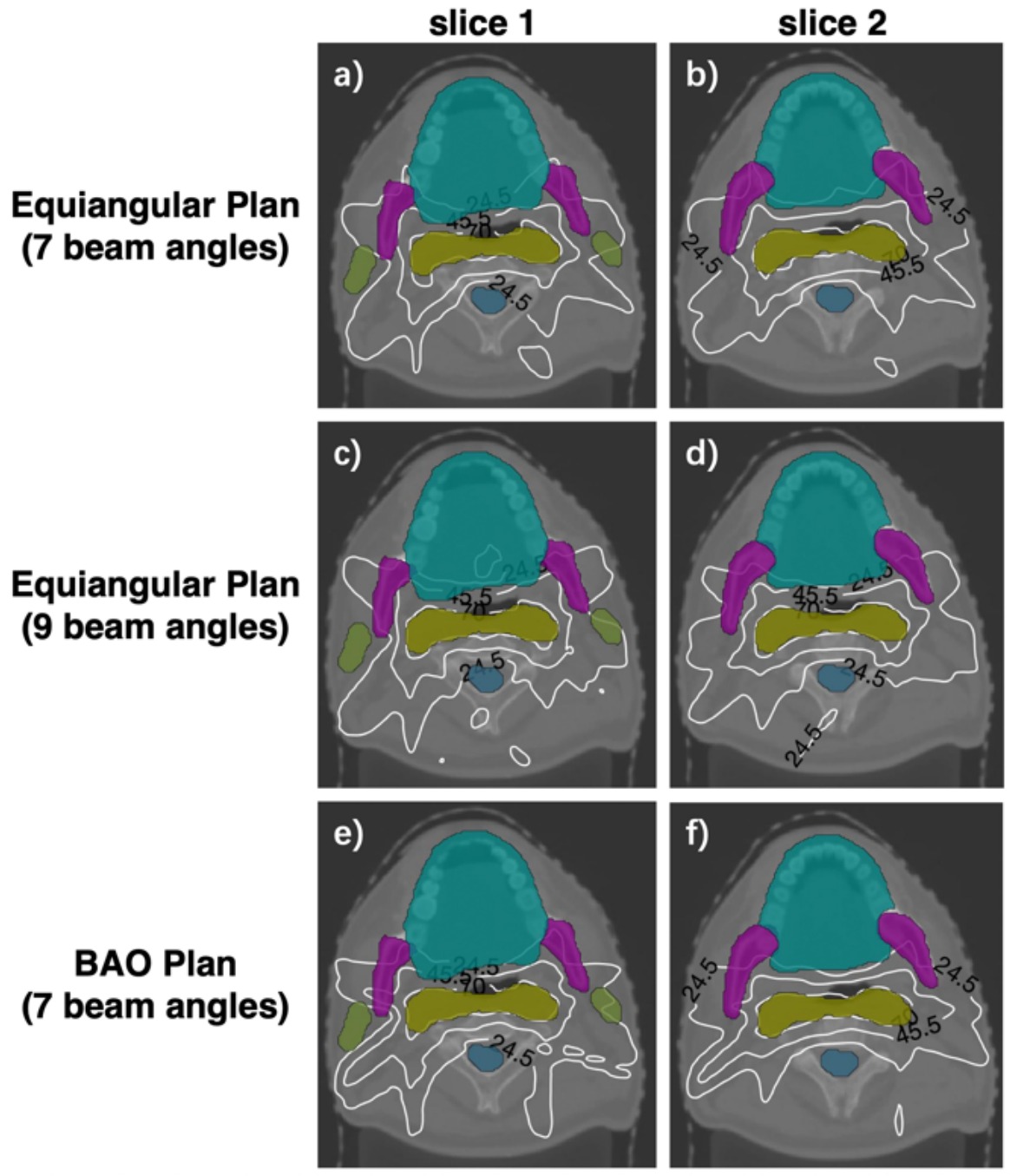

A similar performance of the proposed BAO is observed on the head-and-neck patient, as shown in Figs. 5-8. Fig. 5 compares the Pareto frontiers of the PTV and the OAR dose objectives for the IMRT plans using seven beams generated from the proposed BAO with seven and nine equiangular beams starting at 0°. It should be noticed that the seven beam angles selected by the BAO outperforms not only the equiangular seven beams but also the nine beams, which further confirms the necessity and essentiality of BAO for the complex and non-intuitive scenarios like head-and-neck cases.

The superior performance of the BAO plan for OAR avoidance is visually verified in the comparison of dose distributions in Fig. 6. The algorithm parameters are tuned to obtain the same dose coverage on PTV in all three plans using seven beams selected by BAO (0°, 45°, 63°, 99°, 126°, 261°, 288°), equiangular seven and nine beams. We find that, compared with the prostate patient case, the results of BAO is more conterintuitive on the head-and-neck patient, mainly due to the geometric complexity of PTV and OARs. In this case, our algorithm selects the optimal beams distant from equiangular directions to better adapt the strip-shape of PTV as well as to avoid the OARs. The improved dose sparing on OARs in the BAO plan is seen in the DVH comparison in Fig. 7. The proposed BAO reduces OAR dose by 16.1% and 12.3% from that of the equiangular plans using seven and nine beams, respectively. Table 2 compares the results of the BAO and equiangular plans with the acceptance criteria which indicates that the plans are all clinically acceptable and the superiority of the BAO plan.

FIG. 6: Comparison of dose distributions of different slices on the HN patient with plans using equiangular (a, b) seven beams, (c, d) nine beams and (e, f) seven beams using the proposed BAO, respectively.

Table 2. HN plan objectives and results.

| Regions | Acceptance criteria | BAO | Equiangular (7/9 beams) |
|---|---|---|---|
| PTV | 95%vol≥70 Gy | 70.4 Gy | 70.0/70.8 Gy |
| Brain Stem | max≤54 Gy | 1.9 Gy | 2.2/2.0 Gy |
| Spinal Cord | max≤40 Gy | 19.3 Gy | 18.0/16.6 Gy |
| Parotid Gland | | | |
| Left | mean≤26 Gy | 22.4 Gy | 22.3/23.2 Gy |
| Right | mean≤26 Gy | 12.1 Gy | 14.7/13.4 Gy |
| Mandible | mean≤35 Gy | 18.6 Gy | 20.6/19.3 Gy |
| Oral Cavity | mean≤40 Gy | 11.0 Gy | 14.0/13.4 Gy |

FIG. 7: Comparison of DVH curves for the BAO plan and the equiangular plans.

## 4. Conclusions and discussion

In this paper, we propose a new BAO algorithm to improve fixed-field IMRT. The problem of optimal angle selection is first formulated as $l_1$-norm minimization based on the sparse optimization, and then converted into a highly efficient framework of standard quadratic optimization. On a digital phantom, the proposed BAO successfully finds the theoretically optimal set of beam angles from more than 3, 000, 000 possible combinations. Our algorithm reduces the delivered dose on OARs by 30.53% and 12.3% to 16.1% on a prostate patient and a head-and-neck patient, respectively, compared with that of equiangular IMRT plans with the same PTV dose coverage.

The optimal set of IMRT beam angles varies on different cancer patients. In the era of patient-specific radiation therapy, beam angle selection remains as one of very few procedures missing in the current clinical practice of fixed-field IMRT, mainly due to its high complexity of implementation. Compared with those of existing researches on non-convex or convex BAO algorithms, the main contribution of our work is to show that BAO can be accurately performed using a simple and efficient framework of standard quadratic optimization. As such, the proposed BAO method is practical for improving IMRT dose performance especially on patients with irregular shapes and/or positions of PTV and/or OARs (i.e., head-and-neck patients). Larger dose benefits achieved by BAO are expected on non-conventional IMRT scenarios (e.g., non-coplanar IMRT[17, 34]), where beam angles have additional degrees of freedom. Our algorithm is therefore more attractive in these applications for its mathematical simplicity.

## Acknowledgements

Research reported in this publication was supported by the Fundamental Research Funds for the Central Universities (Grant No. WK2030040089) from Ministry of Education of the P. R. China. The authors have no relevant conflicts of interest to disclose.